\pdfoutput=1

\documentclass[preprint2]{aastex}

\slugcomment{Not to appear in Nonlearned J., 45.}

\shorttitle{Caffau's star}
\shortauthors{Bovino et al.}

\usepackage{natbib}
\usepackage[hyphens]{url}
\newcommand{\arl}[1]{\url{#1}}
\newcommand{\krome}{\textsc{krome} }

\newcommand{\ith}{$i$th }

\begin{document}


\title{The formation of the primitive star SDSS J102915+172927:\\ effect of the dust mass and the grain-size distribution.}

\author{S. Bovino\altaffilmark{1,2}, T. Grassi\altaffilmark{3}, D. R. G. Schleicher\altaffilmark{4}, and R. Banerjee\altaffilmark{1}}
\affil{$^1$Hamburger Sternwarte, Universit\"at Hamburg, Gojenbergsweg 112, 21029 Hamburg, Germany}
\affil{$^2$Kavli Institute for Theoretical Physics, University of California, Santa Barbara, California 93106, USA}
\affil{$^3$Niels Bohr Institute \& Centre for Star and Planet Formation, \O ster Voldgade 5-7, DK-1350 Copenhagen, Denmark}
\affil{$^4$Departamento de Astronom\'ia, Facultad Ciencias F\'isicas y Matem\'aticas, Universidad de Concepci\'on, Av. Esteban Iturra s/n Barrio Universitario, Casilla 160, Concepci\'on, Chile}
\email{stefano.bovino@uni-hamburg.de}






\begin{abstract}
Understanding the formation of the extremely metal poor star SDSS-J102915+172927 is of fundamental importance to improve our knowledge on the transition between the first and second generation of stars in the Universe.
In this paper, we perform three-dimensional cosmological hydrodynamical simulations of dust-enriched halos during the early stages of the collapse process including a detailed treatment of the dust physics. We employ the astrochemistry package \krome coupled with the hydrodynamical code \textsc{enzo} assuming grain size distributions produced by the explosion of core-collapse supernovae of 20 and 35 M$_\odot$ primordial stars which are suitable to reproduce the chemical pattern of the SDSS-J102915+172927 star. We find that the dust mass yield produced from Population III supernovae explosions is the most important factor which drives the thermal evolution and the dynamical properties of the halos. Hence, for the specific distributions relevant in this context, the composition, the dust optical properties, and the size-range have only minor effects on the results due to similar cooling functions. We also show that the critical dust mass to enable fragmentation provided by semi-analytical models should be revised, as we obtain values one order of magnitude larger. This determines the transition from disk fragmentation to a more filamentary fragmentation mode, and suggests that likely more than one single supernova event or efficient dust growth should be invoked to get such a high dust content.
\end{abstract}


\section{Introduction}
The discovery of the extremely metal poor star SDSS J102915+172927 \citep{Caffau2011} has opened the possibility to indirectly probe the conditions of primordial clouds and to explore the environments which have led to the transition between the first (PopIII) and the second generation of stars. The so called ``Caffau's star" is in fact characterized by an extremely low metallicity of 4.5$\times$10$^{-5}$ $Z_\odot$, supposedly formed from a minihalo enriched with dust and metals by the explosion of a massive primordial star \citep{Klessen2012,Schneider2012}. From the observed abundances of Caffau's star it has been inferred that the supernova progenitor should have a mass of \mbox{20-40} M$_\odot$ which during the explosion released a content of dust of  0.01-0.4 M$_\odot$ in the medium  \citep{Schneider2012}. However, the mechanism that likely led to the formation of Caffau's star is far from being well-understood, because the interplay between different processes makes the problem rather intricate, both physically and numerically. Compared to the formation of PopIII stars, where the chemistry is simple, for the second generation of stars the presence of dust, metals, and feedback adds complexity which results in a high computational cost and large uncertainties. Recently, \citet{Smith2015} performed cosmological simulations of a collapsing minihalo enriched by dust and metals obtained from the outcome of a supernova (SN) explosion of a 30~M$_\odot$ PopIII star. 
The outcome of the SN produced a uniform metallicity of $\sim$2$\times$10$^{-5} Z_\odot$, and the collapse led to vigorous fragmentation  and a turbulent density structure. However, their dust model assumes quantities averaged on a standard size distribution\footnote{See \citet{Klessen2014} for an analysis of the inaccuracy and uncertainties of this approach.}.
In addition, the amount of dust and the composition resemble the present-day ISM properties re-scaled by the metallicity, which is known to have a high depletion efficiency $f_\mathrm{dep}$, i.e. a significant amount of metals is locked into dust \citep{Pollack1994}.
On the contrary, realistic models of dust grains formed in PopIII core-collapse SNe reported over the last decade \citep{Todini2001,Nozawa2007,Schneider2006,Schneider2012,
Marassi2015} show very different grain size distributions and compositions, and much lower depletion efficiencies. This is also supported by recent observations of damped Ly-$\alpha$ systems with a metallicity $\sim$10$^{-3} Z_\odot$ \citep{Schady,Zafar2011}.
In previous studies \citet{Dopcke2011,Dopcke2013}, 
starting from an idealized setup, employed a dust model similar to \citet{Smith2015} and explored different metallicities obtaining, through a sink particle algorithm, the mass distribution of the resulting protostellar clumps, which showed a transition from a flat to a peaky distribution with increasing metallicity. 
So far these are the first and the only works where the dust cooling has been included in three-dimensional (3D) simulations of  gravitational collapse of a minihalo, but a comparison between the two studies is difficult as the initial conditions and the numerical code employed are very different.

In a recently submitted paper, \citet{Chiaki2016} employed a comprehensive chemical model coupled with a proper treatment of the dust grains including the grain growth.
They performed simulations of collapsing halos starting from a cosmological initial setup obtained by the outcome of the \citet{Hirano2014} simulations suite. The fragmentation process has been studied assuming an adiabatic collapse for densities above 10$^{16}$~cm$^{-3}$ and they concluded that the fragmentation is strongly related to the metallicity of the gas, but also influenced by the collapse time-scale. 

Even if this study is very detailed and includes a proper model for the dust grains, there are two important questions which have not been addressed: what is the effect of the grain size distribution and of different dust compositions on the thermal evolution and the dynamical properties of the collapsing halo? How does the dust mass yield affect the fragmentation? In this work we provide a quantitative study of the effect of different grain size distributions/compositions on the dynamics of collapsing minihalos enriched by dust for the case of SDSS J102915+172927, starting  from cosmological initial conditions and allowing for a high dynamical resolution.

In the following sections we introduce the different grain size distributions employed in this work, and give the details of our numerical setup. Then we discuss the main results and draw our conclusions.

\section{Numerical methods}\label{sect:distributionscoupling}
\subsection{Chemistry and microphysics}
Chemistry, microphysics, and dust-related processes are treated via the  \krome package\footnote{www.kromepackage.org} \citep{Grassi2014}, which is well suited to accurately and efficiently model chemistry and microphysics in hydrodynamical simulations. The package has been 
employed to study a variety of astrophysical problems, with a wide range of physical and chemical conditions \citep{Bovino2014,Prieto2015,Katz2015,Schleicher2015}, including the formation of supermassive black holes \citep{Latif2014MNRAS,Latif2015} and simulations of star-forming filaments \citep{Seifried2015}.

In this work, we employ a state-of-the-art primordial network, together with the main cooling/heating processes needed to model the collapse of a metal-free minihalo. Specifically, we include: atomic  and Compton cooling as adopted from \citet{cen92}, H$_2$ roto-vibrational cooling updated to \citet{Glover2015}, collisionally induced emission cooling \citep{Grassi2014}, and dust cooling (Sect.~\ref{sec:distributions}). Chemical heating and cooling produced by the formation/destruction of molecular hydrogen is also considered, these include the energy released by the formation of H$_2$ on dust \citep{Hollenbach1979}, that we compute employing the size-dependent rates of \citet{Cazaux2009}\footnote{The \krome setup employed in these simulations and the dust tables will be available on request.}. The choice to not include  metals is dictated by the fact that at this low metallicity the fine-structure metal cooling is negligible as we are well below the critical metallicity ($Z = 10^{-3.5} Z_\odot$) postulated by \citet{Bromm2001}, and recently confirmed with high-resolution numerical simulations \citep{Bovino2014}.

\subsection{Grain size distributions}\label{sec:distributions}
\begin{figure}[!h]
\includegraphics[scale=1.05]{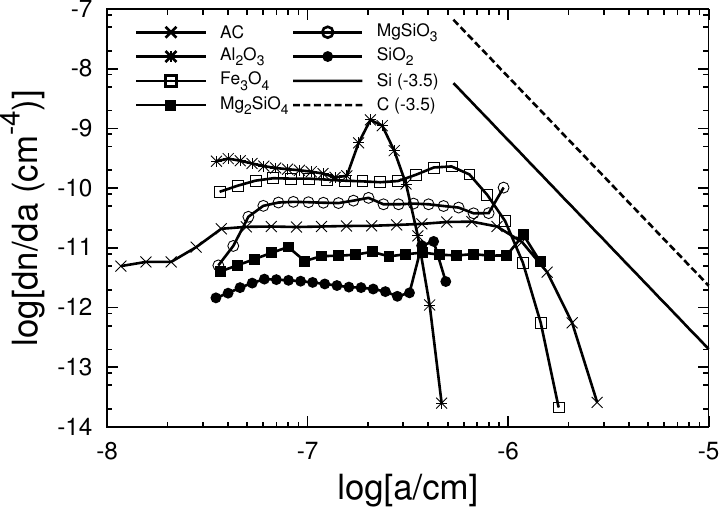}
\caption{Dust grain distributions employed in this work. SN distributions with $f_\mathrm{dep} = 0.0082$ for different dust species, and the standard power-law distributions for C and Si.}\label{fig:figure1_a}
\end{figure}

We explore the effect of the grain size distribution following two different approaches. First, we employ the distribution and dust composition as expected from the outcome of core-collapse SNe explosion \citep{Limongi2012}, which is reprocessed by a nucleation model \citep{Schneider2012, Chiaki2014}. The SN models have been selected to reproduce the average elemental abundances of Caffau's star. These include six different grain species, namely amorphous carbon (AC), alumina (Al$_2$O$_3$), magnetite (Fe$_3$O$_4$), enstatite (MgSiO$_3$), forsterite (Mg$_2$SiO$_4$), and silica (SiO$_2$), with a distribution similar to the one reported in Fig. 2 of \citet{Chiaki2014}, and take into account the passage of a reverse shock which can destroy the initial distribution produced by the SN. We consider here the case of a 35~M$_\odot$ progenitor star (run2 in Table \ref{tab:runs}) where the distribution (reported in Fig. \ref{fig:figure1_a}) is affected by a weak reverse shock (labelled rev1 in Table \ref{tab:runs}) with a depletion factor $f_\mathrm{dep} = 0.0082$, and a case where the progenitor is a primordial star of 20 M$_\odot$ (run3) exposed to a stronger reverse shock (rev2 in Table~\ref{tab:runs}), with a higher depletion factor, $f_\mathrm{dep} = 0.015$. We note that, due to the lack of details in previous papers, it is difficult to retrieve the grain-size distributions for every case. We therefore assume that the grain-size distribution shape is similar and we only change the dust mass yield ($f_\mathrm{dep}$) for the case with 20 M$_\odot$ (run2). As we will show in the next sections this is a good approximation as the shape of the distribution and the grain composition have only a minor impact on the final results. This was also shown by \citet{Ji2014} within a simple one-zone framework.

\citet{Schneider2012} postulated that to have fragmentation at the metallicity of Caffau's star $f_\mathrm{dep}$ should be larger than 0.01.
This threshold has been obtained from simple one-zone models that may not capture the full 3D dynamics and should thus be considered with caution, as also discussed in \citet{Safranek2014MNRAS}. 
With these two distributions we are able to explore values around (run2 and run3) and well above (run4 and run5) the threshold suggested by \citet{Schneider2012} to see under which conditions it has a substantial impact on the density structure. 

A second series of runs is performed including a typical power-law size distribution \mbox{($dn/da\propto a^{-\alpha}$)}, with exponent $\alpha$ = -3.5 \citep{Mathis1977}, where we assume a mix of carbonaceous and silicates resembling the Milky Way (MW) typical composition, with $a_\mathrm{min}$~=~5$\times$10$^{-7}$ and \mbox{$a_\mathrm{max}$ = 10$^{-5}$}~cm as reported in Fig. \ref{fig:figure1_a}. We see that the power-law case has a high mass content ($f_\mathrm{dep} = 0.49$ as provided by \citealt{Pollack1994} and also reported by \citealt{Schneider2012MNRAS}) and spans a smaller size-range. On the other hand, the distribution produced by SN models flattens at smaller radii as an effect of the reverse shock and extends down to very small sizes, of the order of 10$^{-8}$~cm. These differences are relevant to understand the thermodynamics of the system and the results we report in the next sections.
In the case of the power-law (run7), we have to re-scale the solar dust-to-gas ratio by the star's metallicity \mbox{$Z = 4.5\times 10^{-5} Z_\odot$}. As we aim to study the conditions under which Caffau's star was formed we employ the above fixed metallicity value for all the runs presented in this work.

\begin{center}
\begin{figure*}
\includegraphics[scale=0.43]{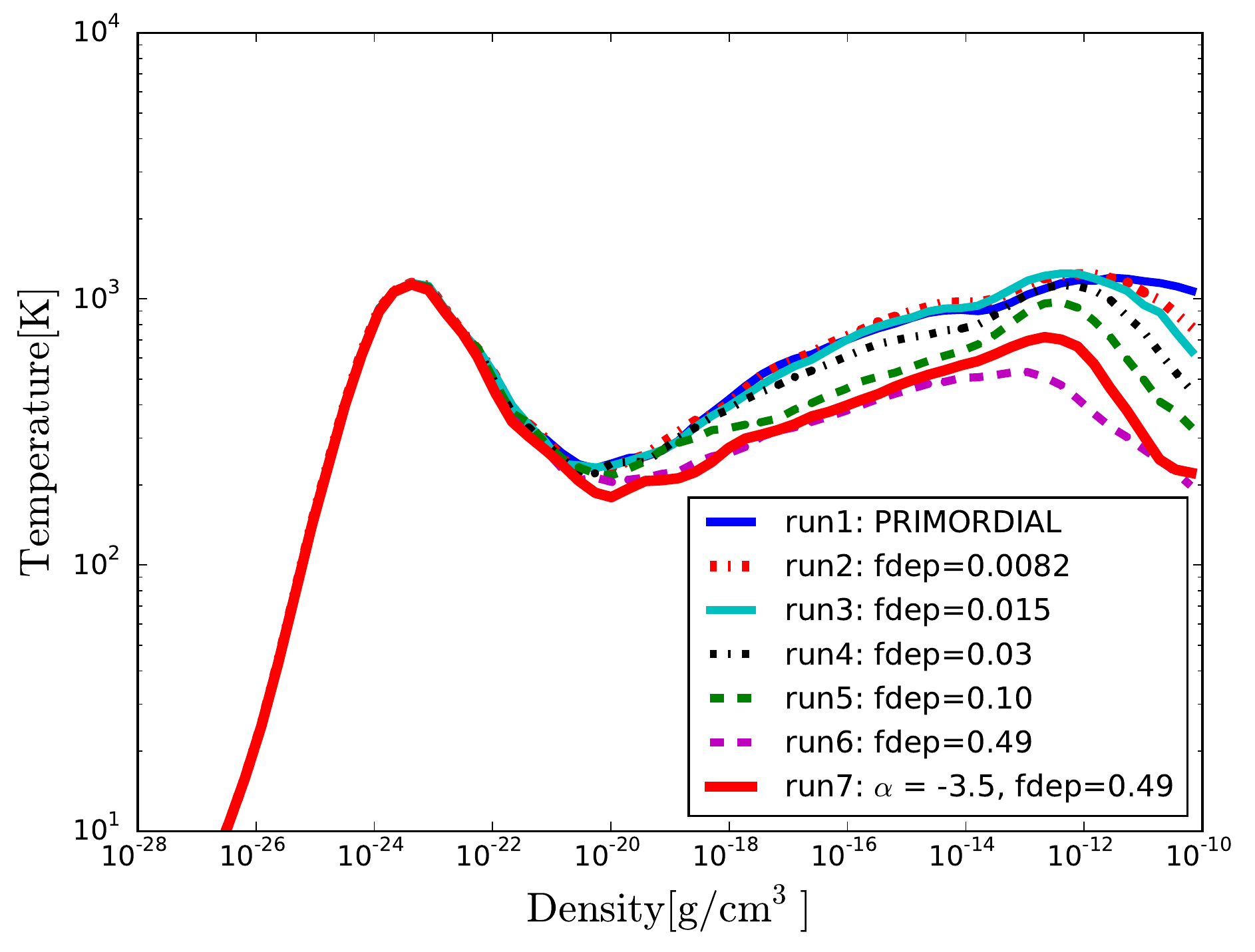}
\includegraphics[scale=0.43]{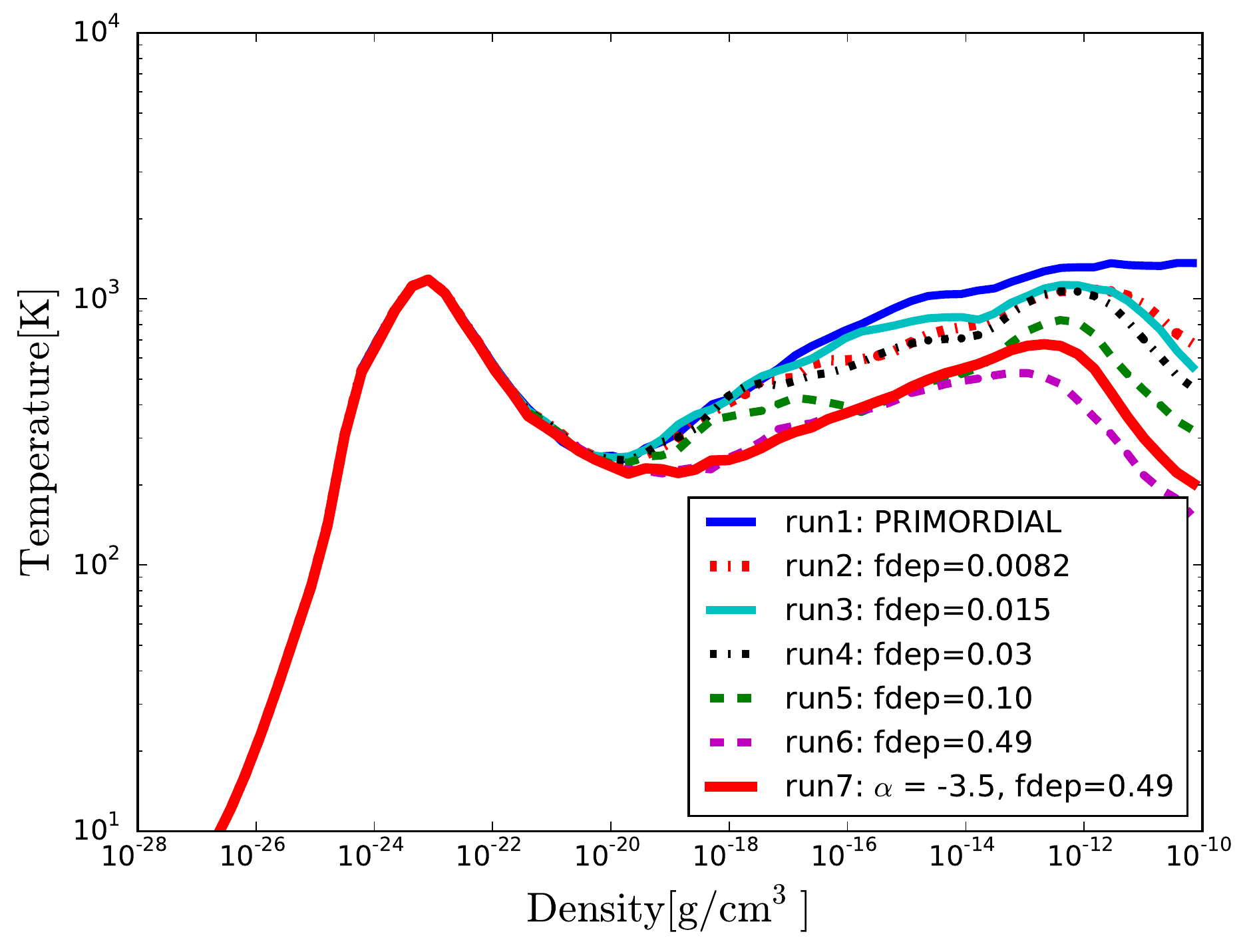}
\caption{Profile of the mass-weighted average temperatures for different size distributions for Halo 1 (left) and Halo 2 (right) taken at the end of the simulation. These are obtained averaging over the radius within a sphere of 300 kpc, starting from the densest point. See text for details.
}\label{fig:figure2}
\end{figure*}
\end{center}

Finally, we also run  a case (run6) where we adopt the 35 M$_\odot$ model, but increase the depletion factor from 0.0082 to 0.49, i.e. assuming the same $f_\mathrm{dep}$ as for the power-law case, to disentangle the effect of the dust mass, composition, and distribution on the dynamics.

The dust physics is employed in \krome via look-up tables obtained by the procedure described in \citet{Grassi2016}, which are only functions of the total density $n_\mathrm{tot}$ and the gas temperature $T$. We assume $N_d = 20$ bins in size per each grain type ($N_t$), which ensure a very good convergence on the results\footnote{Note that an acceptable convergence is already reached for 10 bins.}. 
The total dust cooling is given by summing the contributions over all bins ($N_d\times N_t$) as
\begin{equation}
 	\Lambda_\mathrm{d}=\sum_{i=1}^{N_d\times N_t} \Lambda_{\mathrm{d},i}\,,
\end{equation}
with
\begin{equation}
 \Lambda_{\mathrm{d},i}=2 f v_g n_g \pi L_i\,,
\end{equation}
where $v_g$ is the gas thermal speed, and $f$ takes into account the contributions of species other than protons and is assumed to be equal to 0.5 \citep{Hollenbach1979}. The term $L_i$ is defined as
\begin{equation}\label{eqn:partial_cooling}
 L_i = a_i^2 n_{d,i} k_B\left[T-T_{d,i}\right]\,,
\end{equation}
where $k_B$ is the Boltzmann constant, $T$ the gas temperature, $n_{d,i}$, $a_i$, and $T_{d,i}$, are the dust number density, the grain size, and the dust temperature in the \ith bin, respectively. 
The $N_d\times N_t$ dust temperatures are the roots of the non-linear system of equations
\begin{equation}\label{eq:beta}
 \beta_e(\mathbf{T}_d)\left[\Gamma_{em,i}-\Gamma_{abs,i}\right] = \Lambda_{\mathrm{d},i}\,,
\end{equation}
where \mbox{$i = 1,N_d\times N_t$}, $\Gamma_{em,i}$, and $\Gamma_{abs,i}$ being the radiation emitted and absorbed by a dust grain, and $\beta_{e}$ the escape probability (\citealt{Omukai2000}) defined as
\begin{equation}
 \beta_e(\mathbf{T}_d) = \min\left[1, \,(\tau_g+\tau_d)^{-2}\right]\,.
\end{equation}
\noindent The gas opacity ($\tau_g$) is taken from \citet{Mayer2005}, while the optical constants for every individual dust species ($\tau_d$) are obtained from the refractive indexes reported by the Jena Database (\sloppy\mbox{\url{www.astro.uni-jena.de/Laboratory/OCDB/}}). For the carbonaceous and silicates employed in the power-law distribution, the optical constants are from Bruce Draine's website\footnote{\url{http://www.astro.princeton.edu/~draine/dust/dust.diel.html}} (see \citealt{DraineLee}).  The bulk densities are  from \citet{Nozawa2006}.
We note here that $\beta_e$ includes contributions from all dust bins and their respective temperatures (\mbox{$\mathbf{T}_d = \{T_{d,1},\dots,T_{d,N_d\times N_t}\}$}), thus making the system in Eq. \ref{eq:beta}  non-linear (see \citealt{Grassi2016} for additional details).


\begin{table*}
	\caption{Specifications of the runs presented in
this work: type of distribution, depletion factor $f_\mathrm{dep}$, dust-to-gas mass ratio $\mathcal{D}$, and type of reverse shock destroying the initial dust distribution. The arrows indicate
values above (up) and below (down) the critical $f_\mathrm{dep}$. We employed $Z_\sun = 0.02$ to be consistent with previous estimates of the $f_\mathrm{dep}$ (e.g. \citealt{Pollack1994}). 
The notation 4.9(-9) reads as 4.9$\times$10$^{-9}$. Note that the dust-to-gas mass ratio for the power-law distribution has been evaluated as $\mathcal{D} = \mathcal{D_\sun} \times Z/Z_\sun$ with $\mathcal{D_\sun} = 0.00934$ while, as already discussed in the text, for the SN-like distribution we obtain it from the depletion factor as $\mathcal{D} = f_\mathrm{dep} Z$.}
	\begin{center}
		\begin{tabular}{lllclcc}
			\hline
			\hline
			$\#$ & type &  $f_\mathrm{dep}$ & & $\mathcal{D}$ & $dn/da$ & shock   \\ 
			\hline
			\hline
			run1 & no-dust  & - & - & 0 & -  & -\\ 
			run2 & 35 M$_\odot$ & 0.0082 & $\downarrow$& 7.4(-9) & SN model & rev1$^{\star}$\\
			run3 & 20 M$_\odot$  & 0.015 & $\uparrow$ &1.3(-8) & SN model & rev2$^{\star}$\\
			run4 & 35 M$_\odot$ & 0.030 & $\uparrow$& 2.7(-8) & SN model & \\
			run5 & 35 M$_\odot$ & 0.10 & $\uparrow$& 9.0(-8) & SN model  & \\
			run6 & 35 M$_\odot$ & 0.49 & $\uparrow$ & 4.4(-7) & SN model & \\
			run7 & power-law & 0.49 & $\uparrow$ &  4.2(-7)  & $\alpha = -3.5$ & -\\
		        \hline
		        \hline
		\end{tabular}\\
		\vspace{0.1cm}
		$^{\star}$ see \citet{Chiaki2014} for details.
		\end{center}
	\label{tab:runs}
\end{table*}


\section{Simulation setup}
We select two minihalos of masses 10$^6$~M$_\odot$ and 7$\times$10$^5$ M$_\odot$ and follow the collapse from cosmological initial conditions by employing the hydrodynamical code \textsc{enzo} \citep{ENZO2014}. These minihalos have different properties and start to collapse at different times, $z=22$ and $z=18$, respectively \citep{Latif13b,Bovino2014}.

From $z=99$ to $z=22$ we evolve the halos to reach a central temperature of $\sim$10$^3$~K at a density of $\sim$10$^{-23}$~g~cm$^{-3}$ and then enable the dust machinery. Radiative and mechanical feedback from the previous generation of stars is neglected here and the dust mass fraction is constant during the evolution as we are not considering any process which might form or destroy dust but evaporation. We allow for 29 levels of refinement, which yield a resolution at a sub-AU level, and resolve the Jeans length by 64 cells.
Our refinement strategy is based on over-density, Jeans length and particle mass and is applied during the simulations to ensure that all physical processes are well resolved and the Truelove criterion \citep{Truelove1997,Federrath11} is fulfilled. 
A summary of our runs is reported in Table \ref{tab:runs}. We also include the dust-to-gas mass ratio obtained from the depletion factor $f_\mathrm{dep}$ and the metallicity $Z = 4.5\times 10^{-5} Z_\sun$ as $\mathcal{D} = f_\mathrm{dep} Z$ (see \citealt{Schneider2012}). Note that we employ $Z_\sun$ = 0.02 for consistency with previous work (e.g. \citealt{Pollack1994}) even if this value is now obsolete \citep{Asplund2009}. It is important however to underline that the estimate of $\mathcal{D}$ as well the value of $Z_\sun$ never enter the generation of the tables we have used but we decided to report those to allow for a better comparison with previous investigations.

\section{Results}
In Fig. \ref{fig:figure2} we report the averaged thermal evolution for the runs performed for the two halos. In both cases the general trend is very similar, and clearly shows the effect of dust on the thermodynamics. Depending on the employed grain size distribution, we see a weaker or stronger cooling: the runs with low mass dust yield (run2 to run4) start to cool at densities above 10$^{-12}$~g~cm$^{-3}$ as a result of the dust cooling and reach a minimum temperature of 800~K, 600~K and 400~K for Halo 1, and 700 K, 500 K and  400~K for Halo~2. When we increase the amount of dust the cooling \mbox{(H$_2$ + dust)} is much stronger and the effect is already visible at densities of 10$^{-18}$~g~cm$^{-3}$. The minimum temperature reached both in run6 and run7 is around 200~K and could lead to the formation of clumps with masses \mbox{$<$ 0.01 M$_\odot$}. The same dust content ($f_\mathrm{dep}$ = 0.49) provides similar results regardless of the choice of the distribution. In fact, assuming a power-law  (run7) or a more realistic SN distribution (run6) produces only slight differences of the order of $\sim$50-100~K at densities around $\sim$10$^{-13}-10^{-11}$ g cm$^{-3}$. 

\begin{figure}[!h]
\centering
\includegraphics[scale=0.4]{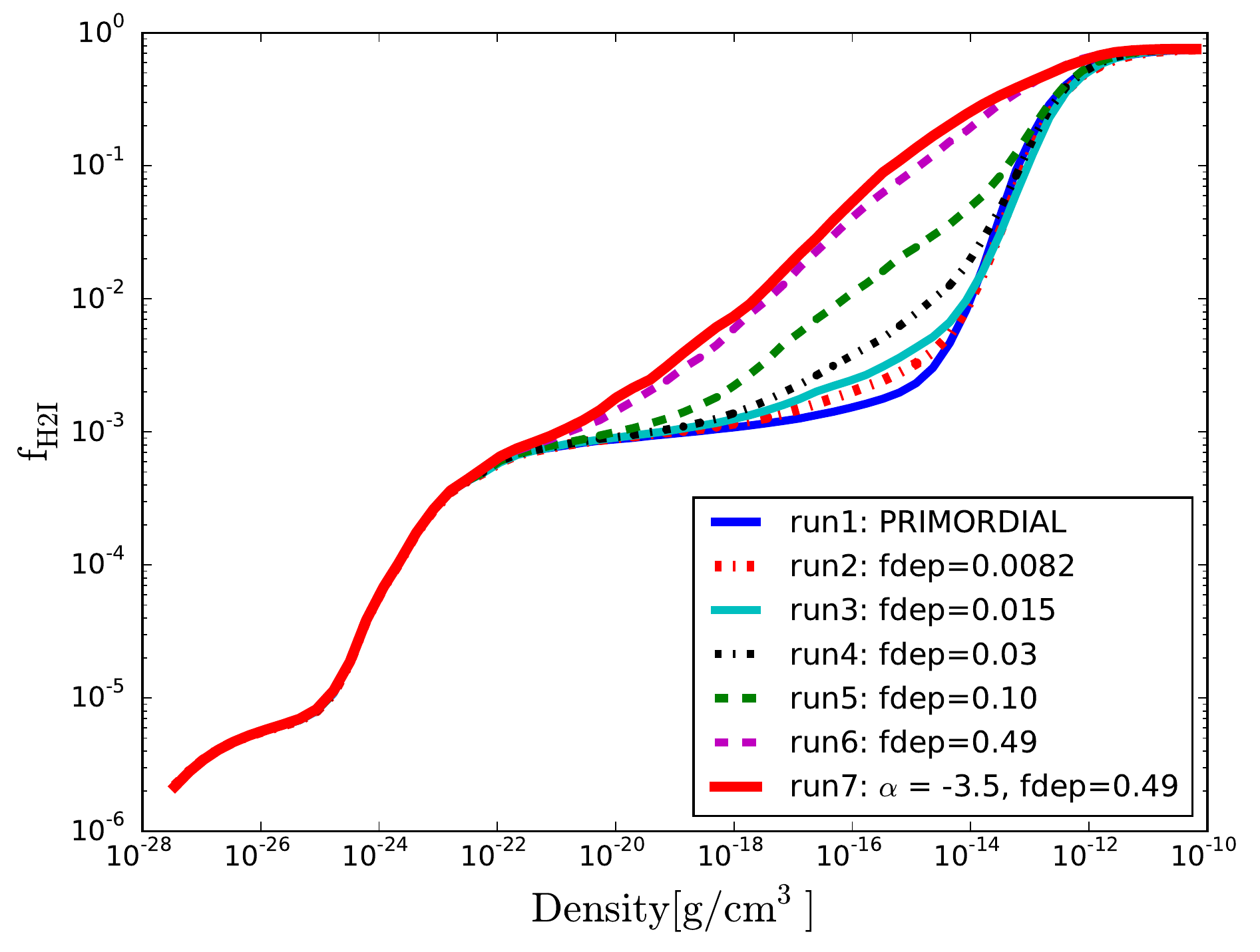}
\caption{Profiles of the mass-weighted average H$_2$ fraction as a function of the density for the different runs discussed in the text. We report only the results for Halo 1.}
\label{fig:h2density}
\end{figure}

In Fig. \ref{fig:h2density} we report the H$_2$ fraction as a function of the density for different runs. As expected from previous studies (e.g. Fig. 4 of \citealt{Omukai2000}) the H$_2$ evolution is strongly affected by the presence of dust, in particular it forms more rapidly while  $f_\mathrm{dep}$ is larger than 0.03. This enhanced H$_2$ fraction results in a slightly higher H$_2$ cooling as discussed in the next section and also reported by \citet{Smith2015}.

\begin{figure}[!h]
\centering
\includegraphics[scale=0.4]{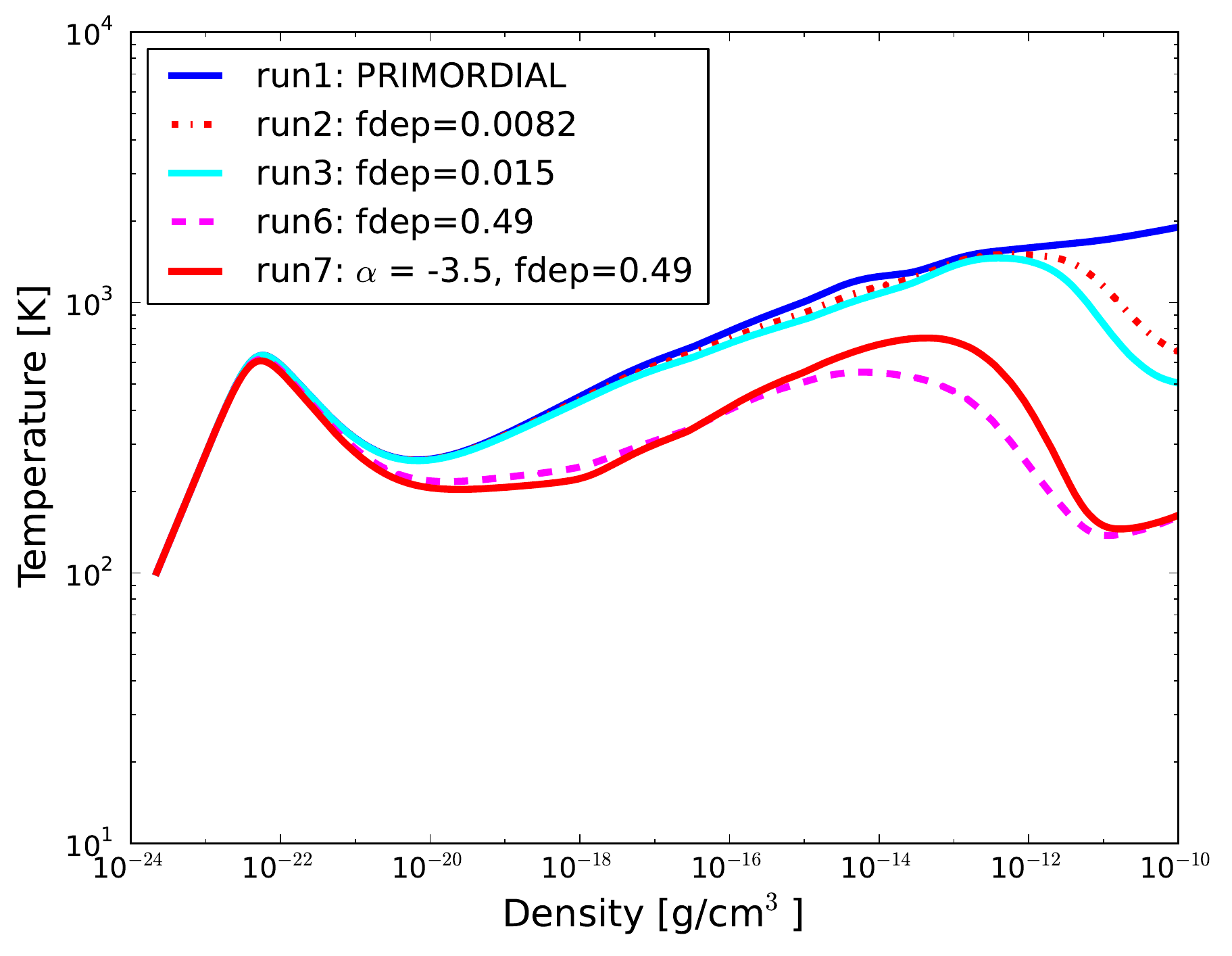}
\caption{Thermal evolution for five specific runs obtained from semi-analytical one-zone models. According to Table \ref{tab:runs} we report run1: primordial case, run2 and run3, i.e. the 35 M$_\odot$ and the 20 M$_\odot$ distributions, with $f_\mathrm{dep}$ 0.0082 and 0.015, respectively, and run6 and run7, that is the SN and the power-law grain-size distributions with the same $f_\mathrm{dep}$ = 0.49.}
\label{fig:onezone}
\end{figure}

\begin{figure}[!h]
\centering
\includegraphics[scale=0.4]{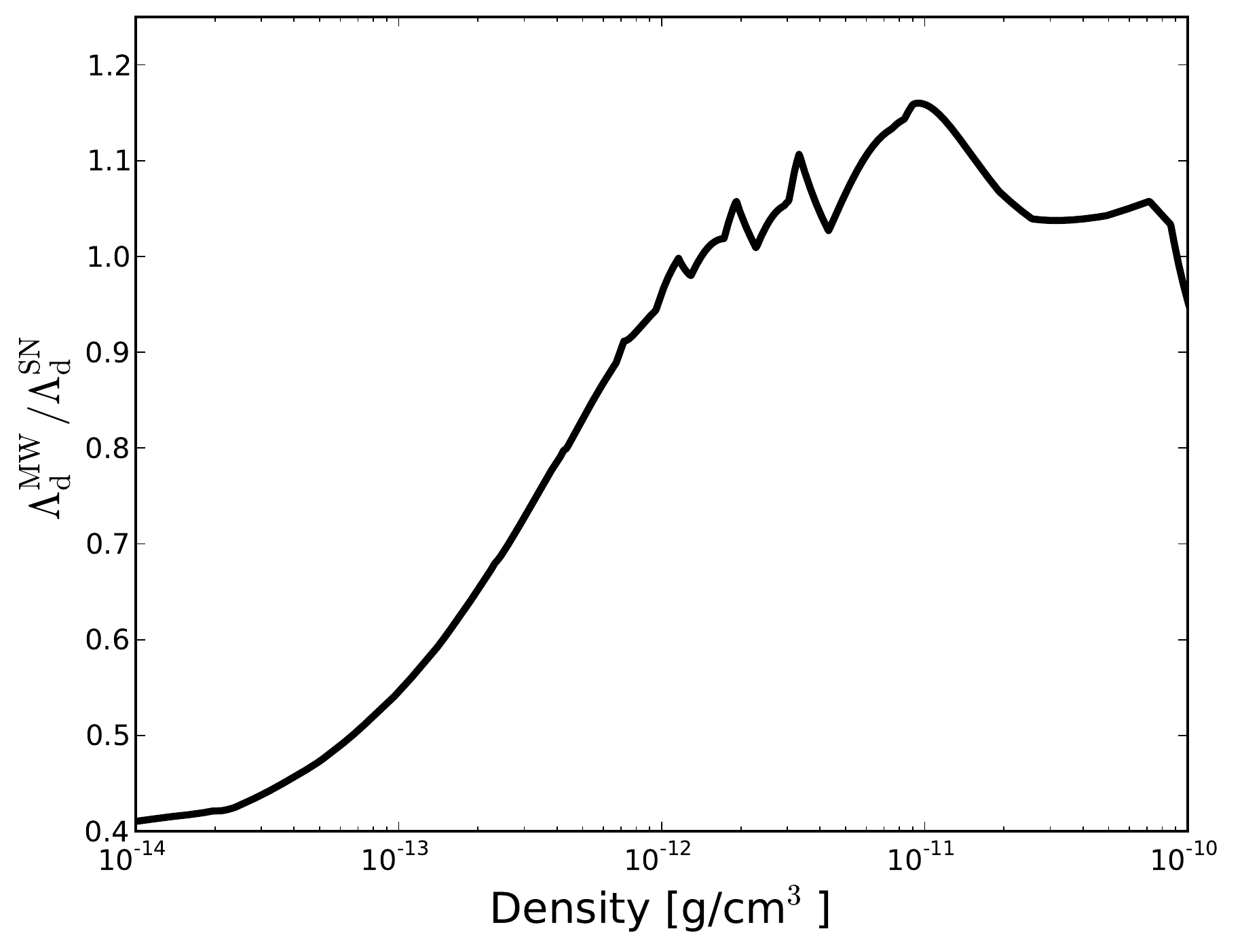}
\caption{Ratio between the cooling $\Lambda_\mathrm{d}^\mathrm{MW}$ provided when employing a Milky Way grain-size distribution, and $\Lambda_\mathrm{d}^\mathrm{SN}$, i.e. the cooling provided when using the grain-size distribution produced by SN models. Only the relevant density region is shown. This ratio has been calculated from the cooling functions obtained from the one-zone models we performed. Note that the dust cooling linearly scales with the dust mass as shown in the Appendix.}
\label{fig:figure6}
\end{figure}

To explore and analyze the differences in the thermodynamics in more detail, we report in Fig.~\ref{fig:onezone} results from one-zone models for the most relevant cases. The overall thermal evolution is in good agreement with the results reported in Fig.~\ref{fig:figure2}. In particular the thermal evolution for run6 and run7, which have the same depletion factor $f_\mathrm{dep}$~=~0.49 but different grain-size distributions, looks very similar to our 3D simulations around densities of 10$^{-14}$-10$^{-12}$~g cm$^{-3}$, i.e. the cooling provided when employing the grain-size distribution produced by SN models is slightly stronger than the one provided by a Milky Way type distribution only at densities between \mbox{10$^{-14}$-10$^{-13}$ g cm$^{-3}$}.
To quantify these differences we plot in Fig.~\ref{fig:figure6} the ratio between the two cooling functions, named $\Lambda_\mathrm{d}^\mathrm{MW}$ and $\Lambda_\mathrm{d}^\mathrm{SN}$ for the Milky Way and the SN-like distributions, respectively. In the region of densities where the dust cooling becomes relevant the ratio varies from 0.4 to 1.2, which means that the SN distribution provides at most two times more cooling compared to the MW-like distribution. This directly reflects the slight difference in the thermal evolution in Fig. \ref{fig:figure2} and explains why we do not see dramatic differences on the final results (thermal and dynamical evolution) even if the two distributions are different in size range, composition, and shape.
To support our numerical results, we provide in the appendix an analytical derivation for the dust cooling which clearly shows, under the given assumptions, that changing the optical properties and the size of the grains has only minor impact on the final dust cooling.


%
%

We should add that there are some differences between the thermal evolution obtained from semi-analytical one-zone models and the three-dimensional hydrodynamical simulations, as a consequence of the fact that in 3D the collapse speed is affected by thermal pressure as well by turbulence and rotation. 
In spite of the agreement in the thermodynamical evolution, we will however show below that the resulting impact on the density structure is different from what was previously assumed in one-zone investigations.

\begin{figure}
\centering
\includegraphics[scale=0.25]{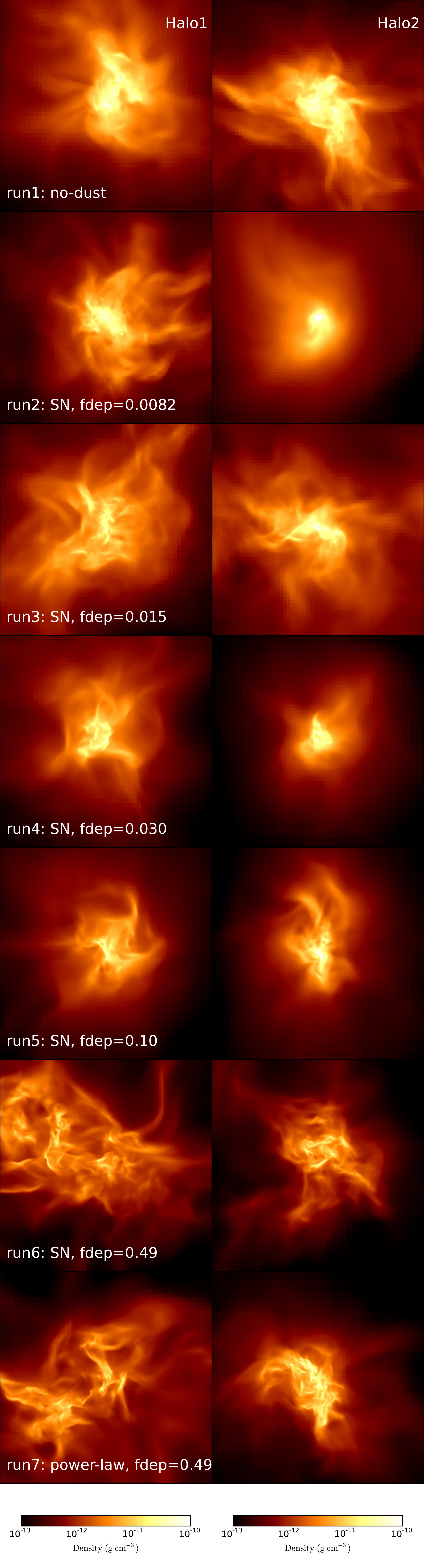}
\caption{Density projections along the x-axis at a scale of 200 AU for the different runs described in Table \ref{tab:runs} and the two halos employed in the simulations.}
\label{fig:figure3}
\end{figure}

\begin{figure}[!h]
\centering
\includegraphics[scale=0.4]{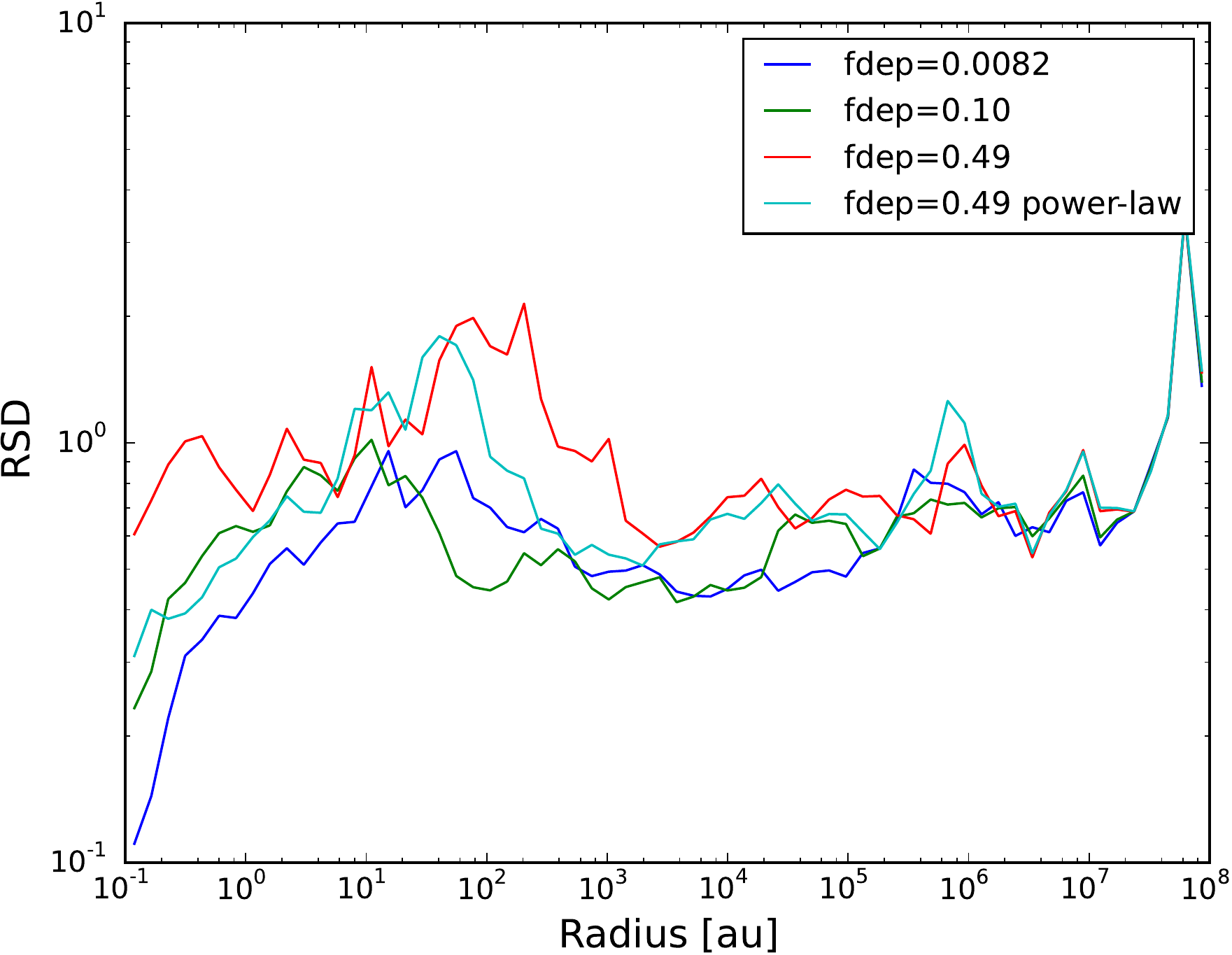}
\caption{Relative standard deviation $\rho_{\mathrm{rms}}/\rho_{\mathrm{mean}}$ as a function of radius for four different runs with Halo 1 as reported in the legend.}
\label{fig:figure4}
\end{figure}

\subsection{Conditions for filaments formation}
To better understand how the differences in the thermodynamics are reflected in the dynamical evolution, we show in Fig.~\ref{fig:figure3} the density projections along the x-axis for the different runs and the two considered halos, taken at a peak density of 3$\times$10$^{-10}$ g cm$^{-3}$. The density structure appears very similar in the two halos even if we note more turbulence in Halo~1. When considering the cases with $f_\mathrm{dep}$ below (run2) or slightly above (run3, run4) the threshold proposed by \citet{Schneider2012}, the collapse proceeds almost monolithically with a turbulent bulky structure which is similar to the primordial case (run1). 

When we analyse the cases with a high depletion factor, i.e. $f_\mathrm{dep} = 0.49$, the gas evolves in a more fragmented  and filamentary structure, reflecting the stronger cooling. The relative standard deviation ($RSD = \rho_{\mathrm{rms}}/\rho_{\mathrm{mean}}$) reported in Fig.~\ref{fig:figure4} confirms that the density dispersion is enhanced when filaments form. The intermediate case with $f_\mathrm{dep} = 0.10$, which is already one order of magnitude larger than the threshold suggested by semi-analytical models, also shows a compact structure. We can then argue that the dust mass threshold lies in the range $0.10< f_\mathrm{dep} \leq 0.49$. A comparison with the thermal evolution further shows that the resulting conditions occur in the regime where $\gamma<1$, i.e. where the temperature decreases in density. This is consistent with previous results by \citet{Peters2012, Peters2014} exploring fragmentation in the presence of a fixed equation of state. 

\begin{figure}[!h]
\includegraphics[scale=0.38]{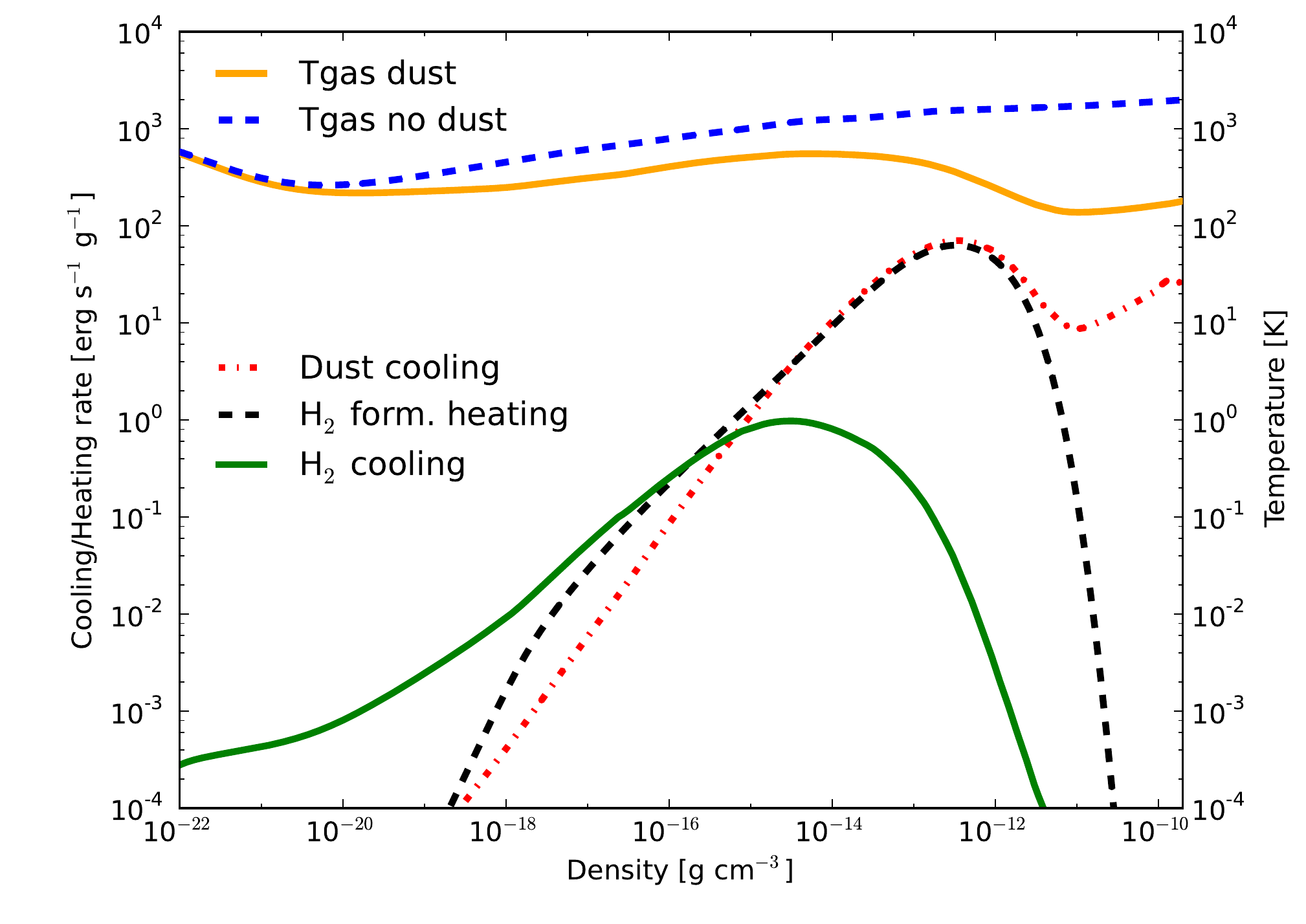}
\caption{Cooling and heating rate as a function of the density for the run with $f_\mathrm{dep}$ = 0.49 (run6 in Table \ref{tab:runs}) . Only the most relevant contributions are shown, namely the H$_2$ line cooling, the dust cooling, and the heating due the formation of H$_2$ both in gas phase and on dust. The thermal evolution is also reported together with the primordial case for reference. Note that the evolution is the same as in Fig. \ref{fig:onezone} but we focus here on higher densities.}\label{fig:figure7}
\end{figure}

To quantitatively assess this point we report in Fig. \ref{fig:figure7} the most relevant cooling/heating rates as a function of the density for run6, i.e. the run where the dust cooling is very strong. As already discussed in previous works (e.g. see \citealt{Omukai2000}) when dust is included both a strong heating due to the catalysis of H$_2$ on grains as well as an increase of cooling at high densities due to collisions between grain dust and gas occur. The H$_2$ abundance, as clear from our Fig. \ref{fig:h2density}, is boosted and so the H$_2$ cooling, at maximum by a factor of two (see also \citealt{Smith2015}). However this enhancement of H$_2$ cooling occurs at intermediate densities and it is exceeded by the H$_2$ formation heating at densities around \mbox{10$^{-16}$ g cm$^{-3}$}.  This enhanced H$_2$ cooling  could have as an effect to boost the formation of HD as reported by \citet{Meece2014}. The only efficient cooling at high densities comes from the dust which is reflected in a sudden drop in the thermal evolution and a $\gamma < 1$. 
As shown by \citet{Peters2012, Peters2014}, the formation of filaments requires an equation of state with $\gamma < 1$. As evident from our Fig. \ref{fig:figure7}, the latter may occur at densities around \mbox{10$^{-22}$-10$^{-20}$ g cm$^{-3}$}, potentially leading to the formation of high-mass clumps, or at densities above 10$^{-14}$ g cm$^{-3}$  which is well within the dust cooling regime. It should also be noted that \citet{Chiaki2016} discussed the possibility that the heating produced by the H$_2$ formation on dust grains could sometimes works against fragmentation.

However at this stage of the simulation we are not able to explore the fragmentation process in detail, and it would be important to evolve the system for longer time, following also the subsequent accretion process. Overall, a thin filamentary structure is expected to favour the formation of low-mass stars as clumps are more easily ejected from the filaments reducing the accretion rate, compared for instance to a thick disk case. In a fragmentation mode based on a thick disk, it is conceivable that more material is available for subsequent accretion, and the resulting clumps may be more massive
\citep{Clark2011,Susa2014,Latif2015,Bromm2016}. However, we cannot rule out the possibility of three-body ejection events, so that the formation of low-mass stars may also be possible in this context (e.g. see \citealt{Ji2014}).

\begin{figure}[!h]
\includegraphics[scale=1.1]{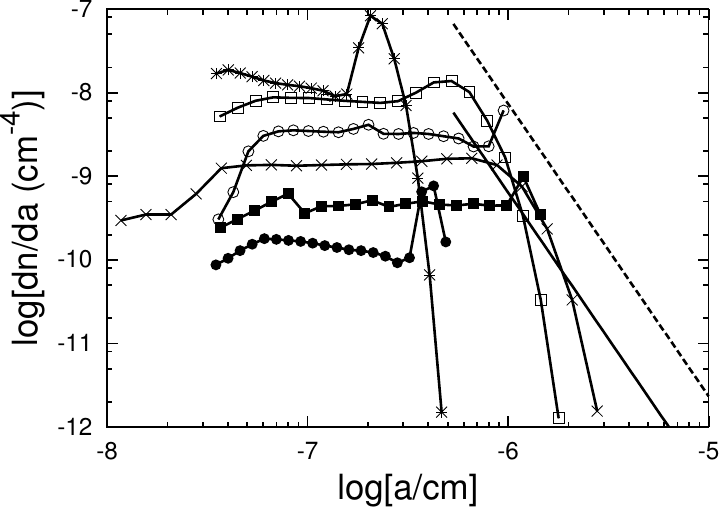}
\caption{Same as figure \ref{fig:figure1_a} but  for the SN distributions where $f_\mathrm{dep}$ has been increased to match the dust mass typical of the power-law distribution, i.e. $f_\mathrm{dep} = 0.49$. Note that the y-scales are different compared to Fig. \ref{fig:figure1_a}.}\label{fig:figure1}
\end{figure}

To better understand the effect of the grain-size distribution we look at the case with an increased dust mass for the SN model distribution (run6), reported in Fig. \ref{fig:figure1}. By increasing $f_\mathrm{dep}$ from 0.0082 to 0.49 we can also catch the effect of the composition/distribution on the thermal evolution and density structure. When comparing the two cases with the same dust content but different distribution and composition (run6 and run7), i.e. different optical properties and size ranges (Fig. \ref{fig:figure1}), we obtain very similar results (see thermal evolution in Fig. \ref{fig:figure2} and \ref{fig:onezone}) and, in particular, similar filamentary density structures (Fig. \ref{fig:figure3}). This crucial result suggests that the grain composition and the shape of the size distribution \textit{do not strongly impact on the evolution of the collapsing halo and its density structure, in particular not as much as the mass yield does}. The latter turns out to be the key physical quantity to understand how Caffau's star was formed.

\section{Conclusions and final remarks}
In this paper we present 3D hydrodynamical calculations of two collapsing minihalos enriched by dust, starting from cosmological initial conditions. We employ the chemistry package \krome \citep{Grassi2014} coupled with the hydrodynamical code \textsc{enzo} including realistic grain size distributions and consistent dust physics. Two different types of grain distributions have been included: i) a standard power-law with composition and dust content similar to our galaxy, i.e. a mix of carbonaceous and silicates, and ii) a more realistic distribution for high-redshift environments \citep{Schneider2012}.
This consists of a sum of log-normal distributions modified by the presence of reverse shocks which flattens the distribution at smaller radii. The two cases allow us to explore different dust compositions and distributions, and assess the effect of the dust physics on the dynamics of the collapsing minihalos. In addition, we explore one case where the outcome of the SN models has been modified by increasing the dust mass yield, and compare this with the results obtained by employing the standard power-law distribution. We found that the dust mass yield, expressed in terms of the depletion factor $f_\mathrm{dep}$, is the key quantity which has a strong impact on the dynamical and thermal evolution of the minihalos.

In fact, when we impose the same dust mass (i.e. $f_\mathrm{dep} = 0.49$) for the standard power-law distribution and the one coming from SN models, these show similar results and dynamical properties.
The composition and the size-range, under the conditions explored here and the grain-size distributions employed, are only producing minor differences because of similar dust cooling power. This is also supported by our analytical derivation in the Appendix.
In addition, our results suggest that a threshold in dust mass exists, as proposed by \citet{Schneider2012}, but we also found that it is at least one order of magnitude larger. Indeed, the formation of filaments depends more on the equation of state parameter $\gamma$ \citep{Peters2012}, rather than local minima in the temperature, as often assumed in one-zone investigations. The fragmentation after the formation of such filaments has been investigated in detail by \citet{Peters2014}.


Our results then suggest that assuming a single supernova event is not enough to provide the necessary dust mass to induce filaments formation. This leads to other scenarios:  i) more than one supernova contributes to the enrichment of the medium and the halo where the star is formed, ii) a weak or no reverse shock should reprocess the dust coming from the SN explosion if a single event is assumed, iii) very efficient grain growth by sticking from the gas phase (in particular on silicates grains) that can increase the dust mass during the collapse should be considered.

However, the latter is rather uncertain as a minimum amount of refractory elements should be present in the gas phase \citep{Chiaki2014}. In addition the grain growth process must be fast enough to increase the depletion factor by orders of magnitudes. \citet{Nozawa2012} provided the conditions under which the grain growth can be efficient, with particular emphasis on the critical refractory elemental abundance needed to growth in function of the initial depletion factor. If we consider the 35 M${_\odot}$ case with $f_\mathrm{dep}$~=~ 0.0082 (run2 in our Table \ref{tab:runs}), to reach a final $f_\mathrm{dep}$ of 0.25  (i.e. above the threshold of 0.1 we found) a \mbox{[Si/H]$\sim$ -4.00} is needed according to Fig.~3 of \citet{Nozawa2012}. This is slightly higher compared to the silicon abundance observed in Caffau's star \mbox{(i.e. [Si/H] = -4.27)} and means that grain growth, for this specific case, should not be very efficient.
When discussing the above scenarios we should also take into account the uncertainties coming from the dust nucleation models.
The formation of dust in SN ejecta in the models included here comes from classical nucleation theory, but other models as the chemical kinetic approach \citep{Cherchneff2009,Biscaro2014} may lead to different results (see e.g. \citealt{Marassi2015}). Also within the same nucleation theory approach, \citet{Nozawa2007} produced different distributions compared to \citet{Schneider2012}  as this strongly depends on the assumptions made in the SN model (see also \citealt{Chiaki2015}). It is then very important to explore other models which can provide a larger amount of dust under different conditions.

A quantitative and detailed comparison with previous 3D calculations \citep{Dopcke2011,Dopcke2013,Meece2014,Smith2015} is difficult to pursue as they employed different dust models and initial setups. 
\citet{Meece2014} aimed at studying the effect of the metallicity and initial conditions (e.g. spin) on the fragmentation process performing simulations of an idealized halo collapse. However, the maximum density reached in their calculations is \mbox{$\sim$10$^{10}$} cm$^{-3}$ while the dust cooling, for the metallicity explored in this work, starts to be efficient at $n > 10^{11}$ cm$^{-3}$ (\mbox{i.e. $\rho > 10^{-13}$ g cm$^{-3}$}) as also discussed by \citealt{Smith2015}. For this reason a direct comparison is not possible.
\citet{Dopcke2013} reported temperatures between 200-500 K for metallicities of \mbox{10$^{-4}$~-~10$^{-5}$ $Z_\odot$}, in line with our results for the power-law distribution, but assuming an average grain size and different optical properties. In \citet{Smith2015} the final temperature reached is slightly higher but the density structure shows similar features as the findings discussed in this work. We further note, that using a realistic depletion factor from the SN models makes a strong difference for the thermodynamics and the dynamical behaviour. We found that the amount of dust in run3-run5 is below the threshold for which dust cooling becomes relevant. On the other hand, assuming a MW-like distribution/composition, with a high depletion factor, as in the work by \citet{Dopcke2011} and \citet{Smith2015}, can lead to an overcooling.

When discussing fragmentation processes it is also important to consider not only the thermodynamics, but also the dynamical processes which can change the picture sketched here. In particular, it is clear that supernovae can change the density structure and thus the initial conditions for star formation. In case of external enrichment in a scenario as modeled by \citet{Smith2015}, the resulting impact may be weaker, as the strength of the supernova shock has already decreased when reaching the new halo, and they also reported that the new material has efficiently mixed with the existing one, providing relatively uniform initial conditions. A more pronounced effect may occur when considering fall-back from supernovae that exploded within the same halo (see \citealt{Ritter2012} and \citealt{Cooke2014} in the context of carbon-enhanced metal poor stars). Such a scenario may have a stronger effect on the remaining mass that is still available for the process of star formation. We however note that in any case the material has to reach at least a Jeans mass if star formation is to occur, so that subsequently a homologous collapse may occur. From the comparison of our one-zone models with the 3D simulations, we expect no strong differences regarding the thermal evolution, even though the outcome with respect to fragmentation may certainly depend on the mass that was initially available. Recently, \citet{Hopkins2015}  introduced the so called ``promoted star formation" mechanism, which can be activated even in minihalos where the total metallicity is low, but a locally high dust mass could exist due to fluctuations induced by the dust dynamics.
This model assumes that the dust distribution and composition in the early Universe is the same as the one observed for our galaxy.

In addition, considering that this star is a very rare object it could also be that it has been formed from specific uncommon dynamical conditions, as for instance halos with very high spin as the ones reported by \citet{Souza2013} and \citet{Stacy2014} even if in the recent calculations by \citet{Meece2014} this seems unlikely to affect the collapse.

To conclude, the results presented in this work provide a relevant step forward to understand the main dust uncertainties and the parameters influencing the dynamical evolution of low-metallicity collapsing halos.

\acknowledgments
\section*{Acknowledgments}
SB, DRGS, and RB thank for funding through the DFG priority program ``The Physics of the Interstellar Medium" (projects BO 4113/1-2, SCHL 1964/1-2, and BA 3706/3-2). TG acknowledges the Centre for Star and Planet Formation funded by the Danish National Research Foundation.  DRGS thanks for funding through Fondecyt regular (project code 1161247) and through the ``Concurso Proyectos Internacionales de Investigaci\'on, Convocatoria 2015'' (project code PII20150171). SB also acknowledges the  kind hospitality of the Kavli Institute for Theoretical Physics (KITP) where this work has been completed, and Andrea Ferrara, Troels Haugb\o lle, and Raffaella Schneider, for fruitful discussions on this topic. We are indebted to Hiroyuki Hirashita for the derivation of the equations reported in the Appendix. This research was supported in part by the National Science Foundation under Grant No. NSF PHY11-25915. The plots of this paper have been obtained by using the $\mathrm{YT}$ tool \citep{Turk2011a}. The simulations have been performed on the German HLRN cluster under the project nip00035.


\appendix
\section{Impact of optical properties, size, and mass on the dust cooling}
If we make some assumptions on the thermal balance equation for the dust-gas interaction (Eq. \ref{eq:beta}), we can disentangle the effect of the size, the mass, and the optical properties on the dust cooling from analytical considerations. First of all let us consider Eq. \ref{eq:beta} in the optically thin regime, i.e.  $\beta_e(\mathbf{T}_d)$ = 1, and neglect the effect of the absorption from the CMB, i.e. all the $\Gamma_{abs,i}$ = 0, considering the interaction gas-grains more efficient than the absorption of UV and visible radiation, which is reasonable in high-density regime. We simply have for the global system that the energy absorbed by the interaction with the gas equates the thermal emission of infrared photons by the grains

\begin{equation}
	\Gamma_{em} = \Lambda_{dust}
\end{equation}

\noindent which is
\begin{equation}\label{eq:solving}
	\int_0^\infty\int_0^\infty 4\pi \kappa_\nu B_\nu(T_d) m_d \varphi(a) da d\nu = \int_0^\infty \pi a^2 n_g v_g 2 k_B (T - T_d) \varphi(a) da\\.
\end{equation}

\noindent  We neglect here the dependency of $T_d$ from the size. The terms in the above equation are described in Table~\ref{tab:definition}.
We can relate the Planck mean opacity to the absorption coefficient $Q_{abs}$ through the following equation

\begin{equation}
	\kappa_\nu = \frac{Q_{abs} \pi a^2}{m_d}\\,
\end{equation}
which becomes

\begin{equation}
	\kappa_\nu = \frac{3}{4}\frac{Q_{abs}}{a \rho_0}\\,
\end{equation}
with $\rho_0$ being the bulk density of a grain (e.g. 2.3 for graphite, see \citealt{Nozawa2006}) and $Q_{abs}$ being the dimensionless absorption coefficient defined as the ratio between the absorption cross section and the surface area of the grain $Q_{abs}~=~\sigma_{abs} / (\pi a^2)$.

Under the assumption of instantaneous thermal equilibrium, the solution of eq. \ref{eq:solving} allows us to obtain the dust temperature. In this specific case we will analyse the right-hand side (RHS) and the left-hand side (LHS) separately. Due to the fact that we neglect the dependency of $T_d$ from the size, we can split the two integrals on the LHS as

\begin{equation}
	4\pi\frac{4}{3}\pi\rho_0\int_0^\infty \kappa_\nu B_\nu(T_d) d\nu\int_0^\infty a^3 \varphi(a) da\\. 
\end{equation}

\noindent After integration over the black-body spectrum and defining the absorption efficiency in the mid-infrared as approximately $\kappa_\nu$ = $\kappa_0 (\nu/\nu_0)^\beta$ we obtain

\begin{equation}
	4\pi\frac{4}{3}\pi\rho_0\kappa_0 C T_d^{4+\beta} \langle a^3\rangle n_d\\,
\end{equation}
where $C$ is a constant factor coming from the integration of the black-body function and has units of \mbox{erg cm$^{-2}$ K$^{-(4 + \beta)}$ s$^{-1}$}, $\langle a^3\rangle$ represents the average over the the grain-size distribution $\varphi (a)$, with $\int_0^\infty \varphi(a) da = n_d$, and $\beta$ usually ranges between 1-2\footnote{The value of the exponent depends upon the nature of the grain, but some observational constraints have been proposed by using the spectral energy distribution of the emitted radiation, see for example \citet{Lagache}.}, with $\nu_0$ being a reference frequency. We can re-write the above equation in function of the dust mass density $\rho_d = n_d  m_d$ to get the final expression for the LHS term

\begin{equation}
	4\pi\frac{4}{3}\pi\rho_0\kappa_0 C T_d^{4+\beta} \langle a^3\rangle \rho_d \frac{3}{4}\frac{1}{\pi\rho_0 \langle a^3\rangle}\\.
\end{equation}
The resulting cooling rate obtained from the LHS term under the assumption of thermal equilibrium is

\begin{equation}\label{eq:lhs}
	\Gamma_{em} \equiv \Lambda_d^{lhs} = 4\pi C \kappa_0 T_d^{4 + \beta} \rho_d\\.
\end{equation}

The RHS after integration over the size distribution is simply

\begin{equation}
	\pi\langle a^2\rangle n_g v_g n_d 2 k_B (T - T_d)\\,
\end{equation}

\noindent and introducing the dust mass density we get the final expression 

\begin{equation}\label{eq:rhs}
	\Lambda_{dust} \equiv \Lambda_d^{rhs} = \frac{3}{2}\frac{\langle a^2\rangle}{\langle a^3\rangle} n_g v_g k_B (T - T_d) \frac{\rho_d}{\rho_0}\\.
\end{equation}

\noindent Equating equation \ref{eq:lhs} and \ref{eq:rhs}, the solution of which is $T_d$, we obtain

\begin{equation}\label{eq:tdustsol0}
	\frac{T - T_d}{T_d^{4 + \beta}} = \frac{8}{3}\frac{\pi\rho_0 C \kappa_0}{n_g v_g k_B}\frac{\langle a^3\rangle}{\langle a^2\rangle}\\.
\end{equation}

\noindent In the limit of very high densities ($n_g > 10^{12}$ cm$^{-3}$), where the dust cooling becomes important, $T_d$ and $T$ are almost coupled, and we can assume $T = T_d + \epsilon$, with $\epsilon$ being a small  number. Equation \ref{eq:tdustsol0} can be approximated to


\begin{equation}\label{eq:tdustsol}
	T_d^{4 + \beta} = \frac{3}{8}\frac{n_g v_g k_B\epsilon}{\pi\rho_0 C \kappa_0}\frac{\langle a^2\rangle}{\langle a^3\rangle}\\.
\end{equation}

\noindent Note that $\epsilon$ slightly changes during the evolution but we can in practice consider it as constant for our analysis. Combining equations \ref{eq:lhs}, \ref{eq:rhs}, and \ref{eq:tdustsol} allows us to make the following considerations given that the LHS and the RHS of Eq. \ref{eq:tdustsol} must be equal:

\begin{itemize}
	\item $T_d$ is independent on the dust mass density $\rho_d$. The only effect of the dust mass is to linearly increase both the LHS (Eq. \ref{eq:lhs}) and the RHS (Eq. \ref{eq:rhs}) terms by the same amount. Generally speaking the dust cooling increases when the dust mass increases. As $\rho_d = \mathcal{D} \rho_g$ = $f_\mathrm{dep} Z/Z_\odot \rho_g$, in terms of depletion factor the dust cooling linearly increases with increasing $f_\mathrm{dep}$ as also shown by our numerical simulations.
	\item Fixing the size $a$: if $\kappa_0$ increases, i.e. the emissivity capacity of a grain increases, $T_d$ only slightly decreases because of the strong dependence  $\sim T_d^{4 + \beta}$. This means that in Eq. \ref{eq:lhs}
	$\kappa_0$ and $T_d^{4 + \beta}$ almost compensate each other and the net effect is a slight change in the cooling. In general  changing the dust composition has a relative small effect on the cooling function.
	\item Fixing the composition, i.e. $\kappa_0$: if $\langle a^3\rangle/\langle a^2\rangle$ in Eq. \ref{eq:tdustsol} increases, i.e. we have an increase in the size $a$, again $T_d$ slightly decreases because physically the volume available to the grain to emit radiation increases ($\Gamma_{em} \propto a^3$) relatively to the dust heating which is proportional to the surface $\propto a^2$. This means that $\langle a^2\rangle/\langle a^3\rangle$ in Eq. \ref{eq:rhs} decreases, but the dust cooling is slightly affected due again to the compensation between the size-term and the dust temperature term.
\end{itemize}

To conclude from our simple analytical analysis, it is clear that the optical properties as well as the size do not strongly affect the final dust cooling unless they change by many orders of magnitudes. The only key quantity which linearly affects the cooling is the dust mass as also shown by our numerical simulations.

{\renewcommand{\arraystretch}{1.4}
\begin{table}
	\begin{tabular}{lccl}
		\hline\hline
		quantities & symbol & formula & units \\
		\hline
		Gas temperature & $T$ & - & K\\
		Boltzmann constant & $k_B$ & - & erg K$^{-1}$\\
		Dust temperature & $T_d$ & - & K\\
		Dust bulk density & $\rho_0$ & - & g cm$^{-3}$\\
		Dust size & $a$ & - & cm\\
		Dust-to-gas mass ratio & $\mathcal{D}$ & $\frac{\rho_d}{\rho_g}$ & - \\
		Dust mass & $m_d$ & $\frac{4}{3}\pi a^3 \rho_0$ & g\\
		Metals mass in gas phase & $m_Z$ & - & g\\
		Depletion factor & $f_\mathrm{dep}$ & $\frac{m_d}{m_d + m_Z}$ & -\\
		Dust number density & $n_d$ & - & cm$^{-3}$\\
		Dust mass density & $\rho_d$ & $n_d m_d$ & g cm$^{-3}$\\
		Dust opacity & $\kappa_\nu$ & $\kappa_0(\frac{\nu}{\nu_0})^\beta$ & cm$^2$ g$^{-1}$\\
		Dust distribution & $\varphi(a)$ & $\frac{dn_d}{da}$ & cm$^{-4}$\\
		Black-body spectrum & $B_\nu$ & $\frac{2 h \nu^3}{c^2} \frac{1}{\exp^{\frac{h\nu}{k_B T_d}}- 1}$ & erg cm$^{-2}$ sr$^{-1}$ s$^{-1}$ Hz$^{-1}$\\
		\hline
		\hline
	\end{tabular}
	\caption{Symbols and mathematical expressions of the quantities used in the appendix. The units are also reported in column 4th.}
	\label{tab:definition}
\end{table}

\end{document}